\documentclass[aps,prd,preprint,groupedaddress]{revtex4-1}
\usepackage{amssymb,amsmath}
\usepackage{dcolumn}% Align table columns on decimal point
\usepackage{bm}% bold math
\usepackage{graphicx}% Include figure files

\newcommand{\tr}{ {\mathrm{tr}\, }}
\newcommand{\Tr}{ {\mathrm{Tr}\, }}

\renewcommand{\det}{ {\mathrm{det} }}

\begin{document}

\preprint{BROWN, INLN}

\title{A Non-Perturbative Gauge-Invariant QCD: Ideal vs. Realistic QCD}


\author{H. M. Fried$^{\dag}$, T. Grandou$^{\ddag}$, Y.-M. Sheu$^{\ddag}$}
\email[]{ymsheu@gmail.com}
%\thanks{}
\affiliation{${}^{\dag}$ {Physics Department, Brown University, Providence, RI 02912, USA} \\ ${}^{\ddag}$ {Universit\'{e} de Nice Sophia-Antipolis, Institut Non Lin$\acute{e}$aire de Nice, UMR 6618 CNRS, 06560 Valbonne, France}}


\date{\today} %{Received: date / Revised version: date}

\begin{abstract}
A basic distinction, long overlooked, between the conventional, "idealistic" formulation of QCD, and a more "realistic" formulation is brought into
focus by a rigorous, non-perturbative, gauge-invariant evaluation of the Schwinger solution for the QCD generating functional in terms of exact
Fradkin representations for the Green's functional $\mathbf{G}_{c}(x,y|A)$ and the vacuum functional $\mathbf{L}[A]$.  The quanta of all (Abelian)
quantized fields may be expected to obey standard quantum-mechanical measurement properties, perfect position dependence at the cost of unknown momenta, and vice-versa, but this is impossible for quarks since they always appear asymptotically in bound states, and their transverse position or momenta can never, in principle, be exactly measured.  Violation of this principle produces an absurdity in the exact evaluation of each and every QCD amplitude.  We here suggest a phenomenological change in the basic QCD Lagrangian, such that a limitation of transverse precision is automatically contained in a proposed `realistic' theory, with the function essential to quark binding into hadrons appearing in the new Lagrangian.  All absurdities in estimates of all "realistic" QCD amplitudes are then removed, and one obtains the possibility of hadron formation by appropriate quark binding potentials, and nucleon scattering and binding by effective, Yukawa-type potentials;
the first of these potentials is constructed, in detail, in a following article.
\end{abstract}

%\keywords{}

\maketitle


\section{\label{SEC:Intro}Introduction}

This is the third paper of a series~\cite{Fried2009_QCD1,Fried2010_QCD2} describing a new, gauge-invariant, non-perturbative formulation of QCD, which provides analytic estimates of any amplitude which sums over all relevant Feynman graphs, including all possible gluon exchanges between appropriate quarks (Q) and antiquarks ($\bar{\mathrm{Q}}$).  All cubic and quartic gluon interactions are included.  Results are expressed in terms of Fradkin's most useful functional representations~\cite{Fradkin1966,HMF2} of the Green's function $\mathbf{G}_{c}(x,y|A) = \langle x | \mathbf{}G_{c}[A] | y \rangle$,
\begin{equation}\label{Eq:1}
\mathbf{G}_{c}[A] = [m + \gamma \cdot (\partial - ig A \cdot \lambda)]^{-1}
\end{equation}

\noindent and the closed-loop, or vacuum functional $\mathbf{L}[A]$,
\begin{equation}\label{Eq:2}
\mathbf{L}[A] = \Tr{\ln{[1 - i g (\gamma \cdot A \cdot \lambda) \, \mathbf{S}_{c}]}}, \quad \mathbf{S}_{c} = \mathbf{G}_{c}[0],
\end{equation}

\noindent where $A_{\mu}^{a}(x)$ represents a given vector potential.  Since the Fradkin representations for these quantities are of Potential Theory origin, they have relatively simple approximations in different physical situations~\cite{HMF3}, especially at high energies where they effectively generate eikonal models.  It should be understood that we are here dealing with the simplest form of "textbook QCD": one type of quark coupled to multicolored gluons.  Flavors and electroweak interactions are to be added later on.

This new formulation provides an exact statement of what may be expected from conventional, or "ideal" QCD, where the quanta of the quark field operators are expected to have the same quantum-mechanical measurement properties as, \emph{e.g.}, electrons (perfect position measurement at the expense of momenta, and vice-versa).  For an important category of QCD processes which involve gluon exchange between quarks, it turns out that this theory is effectively empty, because attention has not been paid to the fact that asymptotic quarks are only found in bound states, and that their transverse coordinates can, in principle, never be exactly determined.  In all previous approximate calculations -- such as those corresponding to gauge-dependent perturbative approximations, or to a subset of Feynman graphs, or to machine calculations wherein gauge-invariance is enforced by breaking up space-time into plaquettes -- this difficulty has not been visible; but for the exact, non-perturbative theory, it becomes clear that quark transverse imprecision must be included as a basic feature of the theory.

The reason for this statement, as well as the relatively simple form of results for each and every QCD amplitude, is due to the novel and unexpected property we call "Effective Locality" (EL), which has been shown in Ref.~\cite{Fried2010_QCD2} to be rigorously true for the exact theory.  In exact QCD, the way in which EL forces the long-held paradigm of Abelian QFT -- where internal interactions are carried by a boson propagator from one space-time point to another -- to change to a new form of locality, will become clear below.

\section{The Fradkin Representations}\label{sec:FradkinRepresentations}

In the first paper on this subject~\cite{Fried2009_QCD1}, we replaced the exact Fradkin representations by their high-energy, eikonal limits; in this paper a more rigorous approach is maintained.  The aim here is to show how the conventional, or "ideal" QCD formulation becomes untenable when treated in this non-perturbative formulation; and then to adopt a more "realistic", though presently phenomenological, version of QCD in which "ideal" troubles disappear, and where the foundation is put in place for automatic quark binding into hadrons~\cite{Fried2011_QCD4}, as well as hadron-hadron interactions of the Yukawa-type~\cite{Fried2011_QCD5}.

For definiteness, we again consider the case of Q-Q or Q-$\bar{\mathrm{Q}}$ scattering, for which the corresponding 4-point function (unsymmetrized, for simplicity of presentation, and in configuration space, before mass-shell amputation) is proportional to
\begin{eqnarray}\label{Eq:3}
\mathcal{N} \int{\mathrm{d}[{\chi}] \, e^{\frac{i}{4} \int{\chi^{2}}}} &\cdot& e^{\mathfrak{D}_{A}} \cdot e^{\frac{i}{2} \int{\chi \cdot \mathbf{F}} + \frac{i}{2} \int{A \cdot \left( \mathbf{D}_{c} \right)^{-1} \cdot A} } \\ \nonumber \quad &\cdot& \left. \mathbf{G}_{c}^{\mathrm{I}}(x_{1}, y_{1} | A) \cdot  \mathbf{G}_{c}^{\mathrm{I\!I}}(x_{2}, y_{2} | A)  \cdot {e^{\mathbf{L}_{\mathrm{c}}[A]}} \right|_{A \rightarrow 0 },
\end{eqnarray}

\noindent where
\begin{equation}
\mathfrak{D}_{A} = - \frac{i}{2} \int {\mathrm{d}}^4x\int {\mathrm{d}}^4y{\frac{\delta}{\delta A^{a}_{\mu}(x)} \,  \mathbf{D}_{c,}{}^{ab}_{\mu \nu}(x-y) \, \frac{\delta}{\delta A^{b}_{\nu}(y)} },
\end{equation}

\noindent $\exp[\mathfrak{D}_{A}]$ is the "linkage operator" which links all pairs of $A_{\mu}^{a}$ to each other and provides the complete panoply of relevant Feynman graphs, $\mathbf{D}_{c,}{}^{ab}_{\mu \nu}$ is the gluon propagator in an arbitrary (relativistic) gauge, $\int{\mathrm{d}[{\chi}]}$ is Halpern's functional integral~\cite{Halpern1977a,Halpern1977b} which allows one to include all cubic and quartic gluon interactions, and $\mathcal{N}$ is a normalization constant such that
\begin{equation*}
\mathcal{N} \int{\mathrm{d}[{\chi}] \, e^{\frac{i}{4} \int{\chi^{2}}}} = 1.
\end{equation*}

\noindent The form of (\ref{Eq:3}), and the mechanism which generates a gauge-invariant result, have been derived, in detail, in Ref.~\cite{Fried2009_QCD1} and \cite{Fried2010_QCD2}, and need not be repeated here.

The Fradkin representations we use are exact variants of those introduced a half-century ago~\cite{Fradkin1966}, which have been re-derived and discussed in detail elsewhere~\cite{HMF2}.  In particular, we find it more convenient to employ a functional integral (FI) representation involving a space-time coordinate $u_{\mu}(s')$ rather than the original, functional differential representation of Fradkin, wherein his $v_{\mu}(s')$ denoted the proper-time-dependent 4-velocity of a real or virtual fermion; the relation between these two quantities is simply:
\begin{equation}\label{Eq:u}
u_{\mu}(s') = \int_{0}^{s'}{\mathrm{d}s'' \, v_{\mu}(s'')},
\end{equation}

\noindent where both $u_{\mu}(s')$ and $v_{\mu}(s')$ are understood to be continuous functions of their proper-time, although no such restriction is placed upon higher derivatives of $v_{\mu}(s')$.  The Fradkin representation of  $\mathbf{G}_{c}(x,y|A)$ is then
\begin{eqnarray}\label{Eq:4}
& & \mathbf{G}_{c}(x,y|A) \\ \nonumber &=&  i \int_{0}^{\infty}{\mathrm{d}s \ e^{-is m^{2}}} \, e^{- \frac{1}{2} \Tr{\ln{\left( 2h \right)}} } \\ \nonumber & & \quad \times \mathcal{N}' \, \int{\mathrm{d}[u]} \, e^{ \frac{i}{2} \int{ u \cdot (2h)^{-1} \cdot u } }  \\ \nonumber & & \quad \times {\left[ m - \gamma_{\mu} \frac{\delta}{\delta u'_{\mu}(s)} \right]} \, \delta^{(4)}(x-y + u(s)) \\ \nonumber & & \quad \times \left( e^{ -ig \int_{0}^{s}{\mathrm{d}s' \, u'_{\mu}(s') \, A_{\mu}^{a}(w) \, \tau^{a}} + g \int_{0}^{s}{\mathrm{d}s' \sigma_{\mu \nu} \, \mathbf{F}_{\mu \nu}^{a}(w) \, \tau^{a}}} \right)_{+},
\end{eqnarray}

\noindent where
\begin{eqnarray*}
w &=& y-u(s'), \\
h(s_{1},s_{2}) &=& s_{2}\theta(s_{1} - s_{2})  + s_{1}\theta(s_{2} - s_{1}) , \\
h^{-1}(s_{1},s_{2}) &=& \frac{\partial}{\partial s_{1}} \frac{\partial}{\partial s_{2}} \delta(s_{1} - s_{2}), \\
\mathcal{N}^{\prime -1} &=& \int{\mathrm{d}[u] \, e^{ \frac{i}{2} \int{ u \cdot (2h)^{-1} \cdot u } }},
\end{eqnarray*}

\noindent and $m$ is the (bare) quark mass.

Explicit $A$-dependence can be extracted from the ordered exponential (OE) of (\ref{Eq:4}), and as in Ref.~\cite{Fried2009_QCD1}, this is illustrated by suppressing the spin-dependent part of the OE. This simplification is done simply for ease of presentation, since conclusions based upon EL are left unchanged~\cite{Fried2010_QCD2}.  We therefore consider the simple OE
\begin{eqnarray*}
& & \left( e^{ -ig \int_{0}^{s}{\mathrm{d}s' \, u'_{\mu}(s') \, A_{\mu}^{a}(y - u(s')) \, \tau^{a}} } \right)_{+} \\ \nonumber &=& \mathcal{N}'' \, \int{\mathrm{d}[\Omega] \, \int{\mathrm{d}[\alpha] \, e^{i\int_{0}^{s}{\alpha \cdot \Omega} } \, \left( e^{ -i \int_{0}^{s}{ \alpha \cdot \tau} } \right)_{+} }} \\ \nonumber & & \times
e^{ -ig \int_{0}^{s}{\mathrm{d}s' \, u'_{\mu}(s') \, A_{\mu}^{a}(y - u(s')) \, \Omega^{a}(s')} },
\end{eqnarray*}

\noindent where $\mathcal{N}''$ is that normalization constant needed for the delta-functional defined by the FI $\int{\mathrm{d}[\Omega]}$.

Because of its definition in terms of $v_{\mu}(s')$, and because of the $\delta^{(4)}(x-y + u(s))$ appearing in (\ref{Eq:4}), one has the "boundary conditions", $u_{\mu}(0)=0$ and $u_{\mu}(s)= - (x-y)_{\mu}$.  With or without that neglected spin-dependence of (\ref{Eq:4}), it is clear that the $A$-dependence of $\mathbf{G}_{c}(x,y|A)$ is not more complicated than Gaussian. This means -- suppressing the $\mathbf{L}[A]$ functional momentarily -- that the needed linkage operations of (\ref{Eq:3}) can be performed exactly, expressing the result in terms of the Halpern and Fradkin FI representations.  As shall become clear immediately, the Fradkin FI representation for $\mathbf{L}[A]$ is also not more complicated than Gaussian: If a simple expansion in powers of $\mathbf{L}$ is adopted, the $A$-dependence of every term in that expansion will not be more complicated than Gaussian either. In brief, the statement that the linkages can be calculated exactly means that the sum over all Feynman graphs is reduced to the summation of a set of FI's~\cite{Fried2010_QCD2}.

After such linkages have been performed, EL appears, which has the extraordinary and simplifying effect of converting the Halpern FI into a set of ordinary integrals.  If the most convenient cluster expansion is used for the linkage operation upon $\exp(\mathbf{L}[A])$, the previous expansion in powers of $\mathbf{L}[A]$ is brought into a much simpler form, and in the end, one need calculate a handful of overlapping proper-time integrals to achieve a reasonable estimate of the desired Physics.  In this paper such detail is not necessary, for to illustrate the distinction between "ideal" and "realistic" QCD, one can simply suppress all $\mathbf{L}[A]$ dependence, using, in effect, a quenched approximation.  In the subsequent papers~\cite{Fried2011_QCD4,Fried2011_QCD5}, explicit examples will be given for the construction of binding and scattering potentials, with and without quenching.

For completeness, we here state the Fradkin representation for $\mathbf{L}[A]$,
\begin{eqnarray}\label{Eq:5}
& & \mathbf{L}[A] \\ \nonumber &=&  - \frac{1}{2} \int_{0}^{\infty}{\frac{\mathrm{d}s}{s} \, e^{-is m^{2}}} \, e^{- \frac{1}{2} \Tr{\ln{(2h)}} } \\ \nonumber && \quad \times \int{\mathrm{d}^{4}x} \, \mathcal{N}^{\prime} \, \int{\mathrm{d}[u]} \, \delta^{(4)}(u(s)) \, e^{ \frac{i}{2} \int{ u \cdot (2h)^{-1} \cdot u} } \\ \nonumber & & \quad \times \left[ \tr{\left( e^{ -ig \int_{0}^{s}{\mathrm{d}s' \, \left[ u'_{\mu}(s') \, A_{\mu}^{a}(w) + i \sigma_{\mu \nu} \, \mathbf{F}_{\mu \nu}^{a}(w) \right] \, \tau^{a}}} \right)_{+}} - 1 \right],
\end{eqnarray}

\noindent which in contrast to (\ref{Eq:4}) contains a Trace over all Dirac and color variables, as well as a four-dimensional $\int{d^{4}x}$ over this closed loop.  It should also be mentioned that $\mathbf{L}[A]$ is rigorously gauge-invariant under the full gauge group of QCD; an original and simple proof may be found in Ref.~\cite{HMF2}.

With the two simplifying assumptions noted above, we write the $A$-dependence of the exponential of the pair of Green's functions of (\ref{Eq:3}), with one $\mathbf{G}_{c}[A]$ representing Q and the other $\bar{\mathrm{Q}}$,
\begin{eqnarray}\label{Eq:6}
\tilde{\mathcal{Q}}_{\mu}^{a} =  &-& g \int_{0}^{s}{\mathrm{d}s_{1} \, u'_{\mu}(s_{1}) \, \Omega^{a}(s_{1}) \, \delta^{(4)}(w - y_{1} + u(s_{1})) } \\ \nonumber &-& g \int_{0}^{\bar{s}}{\mathrm{d}s_{2} \, \bar{u}'_{\mu}(s_{2}) \, \bar{\Omega}^{a}(s_{2}) \, \delta^{(4)}(w - y_{2} + \bar{u}(s_{2})) },
\end{eqnarray}
where the two lines of (\ref{Eq:6}) refer to distinct contributions from each $\mathbf{G}_{c}[A]$. In addition, there is the $A$-dependence from the Halpern FI,
\begin{eqnarray}\label{Eq:chiF}
\int{\mathrm{d}^{4}w \  \chi \cdot \mathbf{F}} = 2 \int{\mathrm{d}^{4}w \, (\partial_{\nu} \chi_{\mu\nu}^{a}) \, A_{\mu}^{a} } + g f^{abc} \int{\mathrm{d}^{4}w \, A_{\mu}^{b} \chi_{\mu\nu}^{a} A_{\nu}^{c} }.
\end{eqnarray}

\noindent The linkage operation to be performed is then
\begin{eqnarray}\label{Eq:7}
\left. e^{-\frac{i}{2} \int{\frac{\delta}{\delta A} \cdot \mathbf{D}_{c} \cdot \frac{\delta}{\delta A} } } \cdot e^{\frac{i}{2}\int{A \cdot \mathcal{K} \cdot A} + i \int{A \cdot (\tilde{\mathcal{Q}} + 2 \partial \chi)}} \right|_{A \rightarrow 0},
\end{eqnarray}

\noindent where $\langle w_{1} | \mathcal{K}_{\mu\nu}^{ab} | w_{2} \rangle = g f^{abc} \langle w_{1} | \chi_{\mu\nu}^{c} | w_{2} \rangle + \langle w_{1} | (\mathbf{D}_{c}^{-1})_{\mu\nu}^{ab} | w_{2} \rangle$.  The operation of (\ref{Eq:7}) is equivalent to a normalized Gaussian FI, and yields
\begin{eqnarray}\label{Eq:8}
e^{-\frac{1}{2} \Tr{\ln{(1 - \mathcal{K} \cdot \mathbf{D}_{c})}} } \cdot e^{\frac{i}{2}\int{ (\tilde{\mathcal{Q}} + 2 \partial \chi) \cdot \mathbf{D}_{c} \cdot \left[ 1 - \mathcal{K} \cdot \mathbf{D}_{c} \right]^{-1} \cdot (\tilde{\mathcal{Q}} + 2 \partial \chi)}} ,
\end{eqnarray}

\noindent using an obvious notation.  And here is where a remarkable mechanism becomes apparent, since
\begin{eqnarray*}
\left[ 1 - \mathcal{K} \cdot \mathbf{D}_{c} \right]^{-1} &=& \left[ 1 - \left(\mathbf{D}_{c}^{-1} + g f \cdot \chi \right) \cdot \mathbf{D}_{c} \right]^{-1} \\ \nonumber &=& \left[ - g (f \cdot \chi) \cdot  \mathbf{D}_{c} \right]^{-1} \\ \nonumber &=& \mathbf{D}_{c}^{-1} \cdot  \left[ - g (f \cdot \chi) \right]^{-1},
\end{eqnarray*}

\noindent so that $\mathbf{D}_{c} \cdot \left[ 1 - \mathcal{K} \cdot \mathbf{D}_{c} \right]^{-1} = - \frac{1}{g} (f \cdot \chi)^{-1} $, and the gauge-dependent propagator has disappeared from the interaction, remaining only in the factor
\begin{eqnarray*}
e^{-\frac{1}{2} \Tr{\ln{(1 - \mathcal{K} \cdot \mathbf{D}_{c})}} } = \left[ \det{(-g f \cdot \chi)} \right]^{-\frac{1}{2}} \cdot \left[ \det{\mathbf{D}_{c}} \right]^{-\frac{1}{2}},
\end{eqnarray*}

\noindent so that the $\left[ \det{\mathbf{D}_{c}} \right]^{-\frac{1}{2}}$ can be safely absorbed into an overall normalization.

We refer the reader to Ref.~\cite{Fried2009_QCD1} for details of exact and approximate evaluations of  $\langle w_{1} | (f \cdot \chi)^{-1} | w_{2} \rangle$.  The important point we here wish to stress comes from the property of locality \begin{eqnarray}\label{Eq:9}
\langle w_{1} | \left. (f \cdot \chi)^{-1} \right|_{\mu\nu}^{ab} | w_{2} \rangle = \left. (f \cdot \chi(w_{1}))^{-1} \right|_{\mu\nu}^{ab} \, \delta^{(4)}(w_{1} - w_{2})
\end{eqnarray}

\noindent We shall call the locality apparent in (\ref{Eq:9}) "Effective Locality" (EL), for it represents a contrast to the way in which ordinary propagators convey information, from one space-time point to another, from $w_{1}$ to $w_{2}$.  Here, (\ref{Eq:9}) is the object which replaces the conventional (perturbative) propagator $\mathbf{D}_{c}(w_{1} - w_{2})$. In Ref.~\cite{Fried2010_QCD2}, this locality is shown to hold for a large family of QCD processes; it simplifies all calculations tremendously, reducing Halpern's FI to a finite set of ordinary integrals, which can be studied numerically.

We illustrate the effects of EL in this paper by making two simplifications, in the interest of clarity:
\begin{enumerate}
  \item Neglect the $\partial \chi$ term compared to $\tilde{\mathcal{Q}}$, which approximation, in view of (\ref{Eq:6}) and (\ref{Eq:chiF}), one would make in a strong-coupling limit. This is just a convenient step for clarity and simplicity of presentation. Most importantly, conclusions are completely independent of this simplification.
  \item The product of the two $\tilde{\mathcal{Q}}$ terms of (\ref{Eq:8}) contains the self-interactions of each particle, as well as the cross-terms, corresponding to the interactions between the particles.  Again, for reasons of simplicity and clarity, we retain only the cross-terms, writing the relevant exponential factor of (\ref{Eq:8}) as
\begin{eqnarray}\label{Eq:10}
& & + \frac{i}{2} g \int{\mathrm{d}^{4}w_{1} \, \int_{0}^{s}{\mathrm{d}s_{1} \, \int_{0}^{\bar{s}}{\mathrm{d}s_{2} \, u'_{\mu}(s_{1}) \, \bar{u}'_{\nu}(s_{2}) \,}}} \\ \nonumber & & \quad \times \, \Omega^{a}(s_{1}) \, \bar{\Omega}^{b}(s_{2}) \,  \left. (f \cdot \chi(w_{1}))^{-1} \right|_{\mu\nu}^{ab} \\ \nonumber & & \quad \times \, \delta^{(4)}(w_{1} - y_{1} + u(s_{1})) \, \delta^{(4)}(w_{1} - y_{2} + \bar{u}(s_{2})),
\end{eqnarray}

\noindent where $u_{\mu}$, $\Omega^{a}$, and $s$ are variables associated with $\mathbf{G}_{c}^{\mathrm{I}}(x_{1}, y_{1}| A)$, whereas $\bar{u}_{\nu}$, $\bar{\Omega}^{b}$, and $\bar{s}$ are with $\mathbf{G}_{c}^{\mathrm{I\!I}}(x_{2}, y_{2}| A)$.  The last line of (\ref{Eq:10}) may be written as
\begin{eqnarray}\label{Eq:11}
\delta^{(4)}(w_{1} - y_{1} + u(s_{1})) \, \delta^{(4)}(y_{1} - y_{2} + \bar{u}(s_{2}) - u(s_{1}))
\end{eqnarray}
\noindent and one sees that, as a consequence of EL, this interaction is associated only with space-time point $w_{1}$.  This means that all the other $w \neq w_{1}$ are irrelevant to the interaction, and as a most important consequence, they are removed, along with their normalization factors, leaving dependence only upon the ordinary integral $\int{\mathrm{d}^{n} \chi(w_{1})}$. Further details are given in Appendix~\ref{AppA}.
\end{enumerate}

The point central to the argument of this article is now being reached, as one seeks to evaluate the support of the second delta-function in (\ref{Eq:11}), which it is convenient to display as a product of delta-functions, in time, longitudinal and transverse coordinates,
\begin{eqnarray}\label{Eq:12}
& & \delta(y_{10} - y_{20} + \bar{u}_{0}(s_{2}) - u_{0}(s_{1}))  \\ \nonumber & & \quad \times \, \delta(y_{1\mathrm{L}} - y_{2\mathrm{L}} + \bar{u}_{\mathrm{L}}(s_{2}) - u_{\mathrm{L}}(s_{1})) \\ \nonumber & & \quad \times \, \delta^{(2)}(\vec{y}_{1\perp} - \vec{y}_{2\perp} + \vec{\bar{u}}_{\perp}(s_{2}) - \vec{u}_{\perp}(s_{1})).
\end{eqnarray}

\noindent In the CM of Q and $\bar{\mathrm{Q}}$, one can choose the origin of each time coordinate as the time of their closest approach; the time difference $y_{10} - y_{20}$ is then always zero.  If the Q and $\bar{\mathrm{Q}}$ are scattering, then $y_{1\mathrm{L}} + y_{2\mathrm{L}} = 0$, since their longitudinal projections are in opposite directions; alternatively, if the Q and $\bar{\mathrm{Q}}$ are bound together, then $y_{1\mathrm{L}} = y_{2\mathrm{L}}$, and their difference vanishes.  Either choice makes no difference at all to the following analysis, and the simplest, second possibility can be adopted, that is, $y_{1\mathrm{L}} - y_{2\mathrm{L}} = 0$.

Consider the time-coordinate delta-function, $\delta(\bar{u}_{0}(s_{2}) - u_{0}(s_{1}))$, which can have a zero argument whenever $\bar{u}_{0}(s_{2})$ and $u_{0}(s_{1})$ coincide.  Assume this happens at a set of points $s_{l}$, so that
\begin{eqnarray}\label{Eq:13}
& & \delta(\bar{u}_{0}(s_{2}) - u_{0}(s_{1}))  \\ \nonumber &=& \sum_{\ell}{\delta(\bar{u}_{0}(s_{\ell}) - u_{0}(s_{1}) + (s_{2} - s_{\ell}) \cdot \bar{u}'_{0}(s_{\ell}) + \cdots)} \\ \nonumber &=& \sum_{\ell}{ \left. \frac{1}{|\bar{u}'_{0}(s_{\ell})|} \, \delta(s_{2} - s_{\ell}) \right|_{\bar{u}_{0}(s_{\ell}) = u_{0}(s_{1})}}.
\end{eqnarray}

\noindent In a similar way, the longitudinal delta-function may be evaluated as
\begin{eqnarray}\label{Eq:14}
\sum_{m}{ \left. \frac{1}{|u'_{\mathrm{L}}(s_{m})|} \, \delta(s_{1} - s_{m}) \right|_{u_{\mathrm{L}}(s_{m}) = \bar{u}_{\mathrm{L}}(s_{2}) \rightarrow  \bar{u}_{\mathrm{L}}(s_{\ell})}},
\end{eqnarray}

\noindent and the product of (\ref{Eq:13}) and (\ref{Eq:14}) as,
\begin{eqnarray}\label{Eq:15}
\sum_{\ell , m}{ \frac{1}{|u'_{\mathrm{L}}(s_{m})|} \, \frac{1}{|\bar{u}'_{0}(s_{\ell})|} \, \delta(s_{1} - s_{m}) \, \delta(s_{2} - s_{\ell}) }
\end{eqnarray}

\noindent under the restrictions $u_{0}(s_{m}) = \bar{u}_{0}(s_{\ell})$ and $u_{\mathrm{L}}(s_{m}) = \bar{u}_{\mathrm{L}}(s_{\ell})$.  Now, $u_{0}$ and $\bar{u}_{0}$, and $u_{\mathrm{L}}$ and $\bar{u}_{\mathrm{L}}$ are continuous but otherwise completely arbitrary functions: The probability that the intersections of $u_{0}(s_{1})$ with $\bar{u}_{0}(s_{2})$ and of $u_{\mathrm{L}}(s_{1})$ with $\bar{u}_{\mathrm{L}}(s_{2})$ occur at exactly the same points is therefore arbitrarily small.  The only place where all four of these continuous functions have the same value is at $s_{1} = s_{2} = 0$, where, by definition of these functions, $u_{\mu}(0) = \bar{u}_{\mu}(0) = 0$, and hence (\ref{Eq:15}) reduces to
\begin{eqnarray}\label{Eq:16}
\frac{1}{|u'_{\mathrm{L}}(0)|} \, \frac{1}{|\bar{u}'_{0}(0)|} \, \delta(s_{1}) \, \delta(s_{2})
\end{eqnarray}

\noindent where, it may be noted, corresponding $u'_{\mathrm{L}}(0)$ and $\bar{u}'_{0}(0)$ appearing in (\ref{Eq:10}) combine with those of (\ref{Eq:16}) to generate the signs, $\epsilon(u'_{\mathrm{L}}(0))$ and $\epsilon(\bar{u}'_{0}(0))$, which may be specified in comparison with those of an eikonal model.

For being crucial to the main argument of this article, it matters to support the heuristic derivation given above with a rigorous mathematical proof.  Since the Wiener functional space~\cite{Lapidus2000} is the most adopted realization of a functional space, the proof will be given in this framework. Here, this space is made out of proper-time parameterized paths: For example, starting from $y_{1\mathrm{L}}$ at $s_1=0$, one will consider paths passing through $u_{\mathrm{L}}(s_{1})$ at proper-time $s_{1} \in \, ]0,s]$; and likewise, paths starting from $y_{2}$ at $s_{2} = 0$ and passing through $\bar{u}_{\mathrm{L}}(s_{2})$ at proper-time $s_{2} \in \, ]0, \bar{s}]$.

%Given this realization of a functional space and $\mu_W$ the associated Wiener measure on it, the condition
%$
%y_{1l}-u_L(s_1)=y_{2l}-\bar{u}_L(s_2)
%$
%expresses the constraint of the second line of (\ref{Eq:12}).

Following the notation convention of Ref.~\cite{Lapidus2000}, $\mathcal{C}^{0,{s}}_0$ is the space ${\mathcal{C}}(]0, s],{\mathbb{R}})$ of continuous functions $u(s_{1})$ from the interval $]0, {s}]$ into ${\mathbb{R}}$, satisfying ${u}(0)=0$; and similar for "barred" quantities.  The $c$-functions $u(s_{1})$ and $\bar{u}(s_{2})$ are here understood to represent either of the 2 possibilities $u_{0}$ and $u_{\mathrm{L}}$.

Endowed with the Wiener measure $m$, $\mathcal{C}^{0,{s}}_0$ is the measurable space, and with the measure $m\!\otimes\! m$, so is the product space $\mathcal{C}^{0,{s}}_0 \!\times \mathcal{C}^{0,{\bar{s}}}_0$ endowed with the topology product, because $u$ and $\bar{u}$ are independent.

Then, we have the following,

\noindent \textit{Theorem:}

For all pair $(s_1,s_2) \in \, ]0,{s}] \, \times \, ]0,\bar{s}]$, one has
\begin{equation}\label{Eq:16A}
m \!\otimes\! m \!\left( \left\{ (u, \bar{u})\in \mathcal{C}^{0,{s}}_0 \!\times \mathcal{C}^{0,{\bar{s}}}_0 \, | \, u(s_1)=\bar{u}(s_2)\right\} \right) = 0
\end{equation}

\noindent whereas
\begin{equation}\label{Eq:16B}
m \!\otimes\! m \!\left( \left\{ (u, \bar{u}) \in \mathcal{C}^{0,{s}}_0 \!\times \mathcal{C}^{0,{\bar{s}}}_0 \, | \, u(0)=\bar{u}(0)=0 \right\} \right) = \left[ m\!\left(\left\{ u \in \mathcal{C}^{0,{s}}_0  \, | \, u(0)=0 \right\} \right) \right]^{2}=1
\end{equation}

\noindent The proof is as follows. Let $\mathrm{B}$ be the set $\left\{ (u, \bar{u}) \in \mathcal{C}^{0,{s}}_0 \!\times \mathcal{C}^{0,{\bar{s}}}_0 \, | \, u(s_1)=\bar{u}(s_2) \right\}$.  One has $\mathrm{B} = \bigcap_{n=1}^{\infty} \mathrm{B}_{n}$, where
\begin{equation}\label{Eq:16C}
\mathrm{B}_{n} = \left\{ (u, \bar{u}) \in \mathcal{C}^{0,{s}}_0 \!\times \mathcal{C}^{0,{\bar{s}}}_0 \, | \, {-\frac{1}{n}} \leq u(s_1)-\bar{u}(s_2) \leq +\frac{1}{n} \right\}
\end{equation}

\noindent Because of the obvious inclusion,
\begin{equation}\label{Eq:16D}
\mathrm{B}_{n+1} \subset \mathrm{B}_{n}, \quad \forall n,
\end{equation}

\noindent  one can write
\begin{equation}\label{Eq:16E}
m \!\otimes\! m(\mathrm{B}) = \lim_{n \rightarrow \infty} m \!\otimes\! m(\mathrm{B}_n).
\end{equation}

\noindent Next, one has~\cite{Lapidus2000}
\begin{eqnarray}\label{Eq:165}
m \!\otimes\! m (\mathrm{B}_n) &=& m \!\otimes\! m \!\left( \left\{ (u, \bar{u}) \in \mathcal{C}^{0,{s}}_0 \!\times \mathcal{C}^{0,{\bar{s}}}_0 \, | \,  u(s_1) - \frac{1}{n} \leq \bar{u}(s_2) \leq u(s_1)+\frac{1}{n} \right\} \right) \\ \nonumber &=& \int_{-\infty}^{+\infty}{\frac{{\mathrm{d}}x}{\sqrt{2\pi s_1}} \, e^{-\frac{x^2}{2s_1}} \, \int_{x-\frac{1}{n}}^{x+\frac{1}{n}} { \frac{{\rm{d}}y}{\sqrt{2\pi s_2}} \, e^{-\frac{y^2}{2s_2}} }} \\ \nonumber &=&  \int_{-\infty}^{+\infty}{\frac{{\rm{d}}x}{\sqrt{2\pi s_1}} \, e^{-\frac{x^2}{2s_1}} \, f_n(x)}.
\end{eqnarray}

\noindent Now, since
\begin{eqnarray}\label{Eq:166}
|\frac{1}{{\sqrt{2\pi s_1}}}\ e^{-\frac{x^2}{2s_1}}\ f_n(x)|\leq \frac{1}{{\sqrt{2\pi s_1}}}\ e^{-\frac{x^2}{2s_1}},
\end{eqnarray}

\noindent the \textit{dominated convergence theorem} can be used so as to take the limit $n\rightarrow \infty$ under the integral and get zero, which establishes (\ref{Eq:16A}), whereas (\ref{Eq:16B}) comes from the normalization procedure of the measure space $\left( m\!\otimes\! m, \mathcal{C}^{0,{s}}_0 \!\times \mathcal{C}^{0,{\bar{s}}}_0 \right)$ into a probability space; the Theorem is then demonstrated.
\par
The product of the first two delta distributions of (\ref{Eq:12}) is therefore proportional to $\delta(s_1)\delta(s_2)$, and further dimensional and symmetry arguments determine the overall multiplicative constant such as in (\ref{Eq:16}). One can observe that this determination of (\ref{Eq:12}) fits the eikonal approximated evaluation of Ref.~\cite{Fried2009_QCD1}.

% commentaire: le raisonnement precedent, menant a (19), se fait sur la base de fonctions de type $C^1$ dont la mesure de Wigner est nulle! Il n'est donc pas conculant dans le cadre de l'espace fonctionnel de Wiener meme s'il conserve quand meme une forte valeur indicative.. ce fait s'etend aux fonctions de type $C^n$ et $C^\infty$.

It is the remaining, transverse delta-function of (\ref{Eq:12}) which is now most relevant, the term $\delta^{(2)}(\vec{y}_{1\perp} - \vec{y}_{2\perp}) = \delta^{(2)}(\vec{b})$, where $\vec{b}$ denotes the impact parameter, or transverse distance between the two particles.  This $\delta^{(2)}(\vec{b})$ is sitting in the exponential of (\ref{Eq:8}), and the question immediately arises as to what meaning it can be assigned.  Depending on its argument, a delta function is either zero or infinite: In the first case this means that there is no interaction, while the second case means that at $\vec{b}=0$, one has an infinite phase factor, suggestive of hard disc scattering~\cite{Schiff}.

The relevant question is therefore why such a delta-function $\delta^{(2)}(\vec{b})$ appears at all.  The answer is that the assumption has earlier been made, and in the conventional, Abelian way, that the $Qs$ and $\bar{Q}s$  may be treated as ordinary particles, whereas it is by now well-known that asymptotic $Qs$ and $\bar{Q}s$ exist only in bound states, and that their transverse coordinates cannot be specified with arbitrary precision. It is therefore unreasonable and unphysical to retain the conventional Abelian practice in which such measurement is assumed possible. This is the interpretation that will be posited here, taking the $\delta^{(2)}(\vec{b})$ outcome as a serious warning that some odd working hypothesis has popped out in exactly this way.

One may wonder why this happens in QCD.  Because QCD possesses EL, which conventional Abelian theories do not. The latter display sums over interconnected propagators, which provide a certain vagueness of position, whereas in the exact non-perturbative QCD, as described above and in Ref.~\cite{Fried2010_QCD2}, one finds the sharp determination of delta-functions corresponding to the EL property, and transverse imprecision must therefore be introduced separately.

In Ref.~\cite{Fried2009_QCD1}, it was suggested that this difficulty be treated in an \emph{ad hoc} phenomenological way, by replacing $\delta^{(2)}(\vec{b})$ by the smoothly varying, effective Gaussian
\begin{eqnarray*}
\varphi(\vec{b})=(2\pi)^{-2} \, \int{\mathrm{d}^{2}\vec{k} \, e^{i \vec{k} \cdot \vec{b} - \frac{k^{2}}{4\mu^{2}}}},
\end{eqnarray*}

\noindent where $\mu$ is a mass parameter on the order of the Q-$\bar{\mathrm{Q}}$ bound state (which we shall call a "model pion"), although we were able to draw the conclusion of that paper without specifying the precise form of $\varphi(\vec{b})$.  In this article, we face this question directly, by first developing a formalism in which transverse quark coordinates cannot be specified, and then showing how this formalism removes all such absurdities, such as that of the exponential factor of $\delta^{(2)}(\vec{b})$ above.  But it must be emphasized that our prescription for such a "realistic" QCD is phenomenological, for there remains to be shown how such an approach could be derived from a more fundamental, operator-field version of QCD, in which transverse imprecision would occur automatically, perhaps in relation to a possible non-commutative geometrical phase of non-perturbative QCD~\cite{Kouneiher2010}.

In a following article~\cite{Fried2011_QCD4}, it will be shown how a specific choice of $\varphi(\vec{b})$, slightly but importantly distinguishable from Gaussian, can serve to define physically reasonable quark binding, and in a subsequent paper~\cite{Fried2011_QCD5}, these techniques will be extended to the construction of nucleon-nucleon scattering and binding potentials.

\section{Towards a possible transverse adapted form of QCD}\label{sec:RealisticQCD}

Perhaps the simplest way of introducing transverse imprecision is to average that part of the QCD Lagrangian dealing with the quark-gluon interaction, so that the transverse position of the color-charge current operator $\bar{\psi} \, \gamma_{\mu} \tau^{a} \, \psi(x)$ should be averaged over a small range by means of an initially unspecified distribution.  One can also demand the same imprecision for the vector current $\bar{\psi} \, \gamma_{\mu} \, \psi(x)$ and scalar density $\bar{\psi} \psi(x)$, but these extra requirements seem to complicate the presentation, to no real advantage, and will not be considered here.

Instead of the conventional quark-gluon contribution to the Lagrangian density,
\begin{eqnarray}\label{Eq:17}
\mathcal{L}_{\mathrm{QG}} = - \bar{\psi} \, [m + \gamma_{\mu} \, (\partial_{\mu} - i g A_{\mu}^{a} \tau^{a})] \, \psi,
\end{eqnarray}

\noindent in which all field operators occur at the same space-time point, and for which gauge invariance under the standard QCD gauge transformations is obvious, we now adopt a local -- in time and longitudinal position -- but non-local in its transverse coordinates replacement,
\begin{eqnarray}\label{Eq:18}
\mathcal{L}_{\mathrm{QG}} &=& - \int{\mathrm{d}^{2} \vec{x}'_{\perp} \, \mathfrak{a}(\vec{x}_{\perp} - \vec{x}'_{\perp}) } \\ \nonumber & & \quad \times \bar{\psi}(x') \, [m + \gamma_{\mu} \, (\partial'_{\mu} - i g A_{\mu}^{a}(x) \tau^{a})] \, \psi(x'),
\end{eqnarray}

\noindent where the transverse imprecision function (TIF) $\mathfrak{a}(\vec{x}_{\perp} - {\vec{x}}'_{\perp})$ is a real, symmetric function of its arguments, of significant value only for distances on the order of the inverse of the pion mass, $x'_{\mu} = (x_{0}, x_{\mathrm{L}}; \vec{x}'_{\perp})$, $\partial'_{\mu} = (\frac{1}{i} \frac{\partial}{\partial x_{0}}, \frac{\partial}{\partial x_{\mathrm{L}}}; \frac{\partial}{\partial \vec{x}'_{\perp}})$, and $A_{\mu}^{a}(x)$ is left untouched.  In this formulation, rigorous local gauge-invariance is suppressed for the underlying quark fields, whose quanta have unmeasurable transverse positions, but the hadrons all constructed from these quanta will nevertheless be proper singlets under SU(3).

One notes that in the contribution of (\ref{Eq:18}) to its part of the Action operator, $\int{\mathrm{d}^{4}x \, \mathcal{L}_{\mathrm{QG}} }$, the $\vec{x}_{\perp}$ and $\vec{x}'_{\perp}$ coordinates can be interchanged, which yields an equivalent form in which every $A_{\mu}^{a}(x)$ of the original (\ref{Eq:17}) is replaced by $\int{\mathrm{d}^{2}\vec{x}'_{\perp} \, \mathfrak{a}(\vec{x}_{\perp} - \vec{x}'_{\perp}) \, A_{\mu}^{a}(x') }$. This interchange allows a very simple extraction of all such transverse imprecision, since both delta-functions of (\ref{Eq:10}) will now be replaced by
\begin{eqnarray}\label{Eq:19}
& & \int{\mathrm{d}^{2} {\vec{y}}_{1\perp}^{\, '} \, \mathfrak{a}(\vec{y}_{1\perp} - \vec{y}_{1\perp}^{\, '}) \, \int{\mathrm{d}^{2}\vec{y}^{\, '}_{2\perp} \, \mathfrak{a}(\vec{y}_{2\perp} - \vec{y}^{\, '}_{2\perp}) \, }}\\ \nonumber & & \quad \times  \delta^{(4)}(w_{1} - y'_{1} + u(s_{1})) \,  \delta^{(4)}(w_{1} - y'_{2} + \bar{u}(s_{2})).
\end{eqnarray}

%% additions 4/9/2011

\noindent This can be seen in the simplest way by noting that the presence of a TIF $\mathfrak{a}(\vec{y}_{\perp} - \vec{y}^{\, '}_{\perp})$ shall mean that the difference $|\vec{y}_{\perp} - \vec{y}^{\, '}_{\perp}|$ is effectively bounded by $1 / {\mu}$, where $\mu$ appears in the definition of $\varphi(b)$ below, and where, as shown in Ref.~\cite{Fried2011_QCD4}, $\mu$ is on the order of the pion mass, $m_{\pi}$.  Hence a negligible error is made by replacing the transverse coordinate $w = y'_{\perp}$ by $y_{\perp}$.  A more precise justification appears in Appendix~\ref{AppB}.  It should be noted that this approximation concerning the utility of EL can be justified in the same way as described in Appendix~\ref{AppB} for both the construction of the quark binding potential and the nucleon-nucleon scattering and binding potential (the exchange of a gluon bundle across a closed quark loop apparently involves a new set of $\int{\mathrm{d}^{n}\chi}$ integrals, and will be discussed elsewhere).

The first delta-function of (\ref{Eq:19}) defines the argument $w_{1}$ of $\chi(w_{1})$, and we again observe that the final output of the Halpern FI will be an ordinary integral $\int{\mathrm{d}^{n}\chi}$, independent of the choice of $w_{1}$.  The second delta-function of (\ref{Eq:19}) now involves the $\mathfrak{a}$-dependence, generating in place of the $\delta^{(2)}(\vec{y}_{1\perp} - \vec{y}_{2\perp})$ which follows from (\ref{Eq:16}), the combination
\begin{eqnarray}\label{mu}
& & \int{\mathrm{d}^{2}\vec{y}^{\, '}_{1\perp} \, \int{\mathrm{d}^{2}\vec{y}^{\, '}_{2\perp} \, \mathfrak{a}(\vec{y}_{1\perp} - \vec{y}^{\, '}_{1\perp}) \, \mathfrak{a}(\vec{y}_{2\perp} - \vec{y}^{\, '}_{2\perp}) }} \, \delta^{(2)}(\vec{y}^{\, '}_{1\perp} - \vec{y}^{\, '}_{2\perp}) \\ \nonumber &=& \int{\mathrm{d}^{2}\vec{y}^{\, '}_{\perp} \, \mathfrak{a}(\vec{y}_{1\perp} - \vec{y}^{\, '}_{\perp}) \, \mathfrak{a}(\vec{y}_{2\perp} - \vec{y}^{\, '}_{\perp})}.
\end{eqnarray}

\noindent Inserting 2-dimensional Fourier transforms of each
\begin{eqnarray*}
\mathfrak{a}(\vec{y}_{\perp} - \vec{y}^{\, '}_{\perp}) = \int{\frac{\mathrm{d}^{2}\vec{k}}{(2\pi)^{2}} \, e^{i\vec{k} \cdot (\vec{y}_{\perp} - \vec{y}^{\, '}_{\perp})} \, \tilde{\mathfrak{a}}(\vec{k}_{\perp})},
\end{eqnarray*}

\noindent the combination (\ref{mu}) becomes
\begin{eqnarray}\label{Eq:20}
\int{\frac{\mathrm{d}^{2}\vec{k}}{(2\pi)^{2}} \, \tilde{\mathfrak{a}}(\vec{k}) \, \tilde{\mathfrak{a}}(-\vec{k}) \, e^{i\vec{k} \cdot (\vec{y}_{1\perp} - \vec{y}_{2\perp})} }.
\end{eqnarray}

\noindent From its definition, $\mathfrak{a}$ is real, and hence (\ref{Eq:20}) becomes
\begin{eqnarray}\label{Eq:21}
\int{\frac{\mathrm{d}^{2}\vec{k}}{(2\pi)^{2}} \, \, e^{i\vec{k} \cdot \vec{b}} \, \left|\tilde{\mathfrak{a}}(\vec{k})\right|^{2} } \equiv \varphi(\vec{b}),
\end{eqnarray}

\noindent which provides the definition of $\varphi(\vec{b})$.  Note that while no restriction has been placed on the form of $\mathfrak{a}$ other than that it is real, from (\ref{Eq:21}), $\varphi$ turns out to be independent of the direction of $\vec{b}$, that is, $\varphi(\vec{b})=\varphi(b)$.

One might expect that an intuitive choice such as $\varphi(b) \sim e^{-\mu^{2} b^{2}}$ would suffice, but that is not true.  By trial and error it is found that a slight change to
\begin{eqnarray}\label{Levy}
\varphi(b) \sim e^{-(\mu b)^{2+\xi}}
\end{eqnarray}

\noindent with $\xi$ real, positive, and small, $\xi<1$, is most appropriate and leads to a binding potential of form $V_{B}(r)=\xi \mu (\mu r)^{1+\xi}$, for small $\xi$~\cite{Fried2011_QCD4}.  For the time being, the interpretation of the parameter $\xi$ is quite enigmatic. There are steps in the derived expression for $V_{B}(r)$ where $\xi$ may be neglected compared to unity; but $\xi$ cannot be set equal to zero.  And $\xi$ is not a coupling constant, for even when small, its effect on the Q-$\bar{\mathrm{Q}}$ binding is large.  It would seem that, in some sense, $\xi$ has something to do with the transverse correlations between bound quarks and/or antiquarks.  In this respect it is worth noticing that (\ref{Levy}) is a part of a Levy-flight probability distribution~\cite{levy}. At this point, a part of the enigma is precisely how such correlations would emerge from the assumed transverse imprecision of the quark field operators.

One other point deserves mention concerning the scale change used when $\int{\mathrm{d}^{4}w \, \chi^{2}(w)}$ in the exponent of the Halpern representation is broken up into small cells of volume $(\delta)^4$:
\begin{equation*}
\int d^4w \chi^2(w) \rightarrow (\delta)^4  \sum_ i  \chi^2_i\ , \ \ \ \  \chi_i \equiv \chi(w_i)
\end{equation*}

\noindent Upon rescaling, $\chi_i \rightarrow (\delta)^{-4} \, \chi'_i$, and re-expressing all interactions in terms of $\chi'$,  there appears in (\ref{Eq:10}) the factor $(\delta)^{2} \, \varphi(b)$.  In Ref.~\cite{Fried2009_QCD1}, where transverse imprecision was treated in an {\it{ad hoc}} way, the size of $\delta$ was taken to be $M^{-1}$, where $M$ corresponded to a very large energy associated with the eikonal limit.  Here, we ask the more physical question of just how small that $\delta$ may be chosen in the light of an actual measurement, and we let Quantum Physics provide the answer:  That contribution to the $\delta$ corresponding to a time separation should be chosen as $1/E$, that corresponding to a (CM) longitudinal coordinate should be $1/p_{\mathrm{L}} \simeq 1/E$, while that corresponding to each of the transverse coordinates should be $1/\mu$; and hence $\delta^{4} \sim 1/(E \, \mu)^{2}$.  An alternate way of expressing this is that, starting from arbitrarily small separations in each coordinate, we average each $\chi_{n}$ variable over a physically meaningful distance, and call that average the $\chi_{n}$ contained in the volume $(\delta)^{4}$.

Finally, we note that $\varphi(b)$ replaces the $\delta^{(2)}(b)$ of the conventional, `ideal' formulation of QCD, with obvious normalization: $\int{\mathrm{d}^{2}b \, \delta^{(2)}(b)} = 1$.  If we adopt the same normalization for $\varphi(b) = \varphi(0) \exp{[-(\mu \, b)^{2+\xi}]}$, then it is easy to see that, with $\delta^{2} = 1/(E \, \mu)$, $ \varphi(b)$ becomes
\begin{eqnarray}
\frac{\mu^{2}}{\pi} \, \frac{1+\xi/2}{\Gamma(\frac{1}{1+\xi/2})} \, e^{-(\mu b)^{2+\xi}} \simeq \frac{\mu^{2}}{\pi} \, e^{-(\mu b)^{2+\xi}}, \quad \xi \ll 1.
\end{eqnarray}

\section{Bundle Diagrams}\label{sec:BuddleDiagrams}

\begin{figure}
\includegraphics[height=35mm]{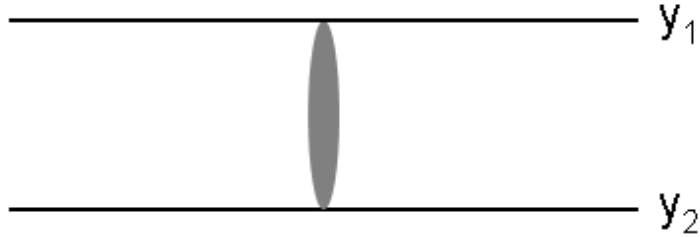}%
\caption{\label{Fig:1}A gluon bundle exchanged between quarks $\mathrm{I}$ and $\mathrm{I\!I}$}
\end{figure}

In the above example of quark and/or antiquark scattering, where the infinite number of exchanged gluons appears to originate and end at a single space-time point on a quark/antiquark line, it may be helpful to introduce the concept of an exchanged `gluon bundle', as in Figure~{\ref{Fig:1}}.  Because of the four-dimensional delta function $\delta^{(4)}(y'_{1} - y'_{2} - u(s_{1}) + \bar{u}(s_{2}))$, arising from the product of the pair of delta functions of (\ref{Eq:19}), and of the subsequent analysis which produces (\ref{Eq:20}), the transverse separation $\vec{b} = \vec{y}_{1\perp} - \vec{y}_{2\perp}$ satisfies the distribution of (\ref{Eq:21}).  But the argument $w$ in $(f \cdot \chi(w))^{-1}$ is given by $w = w_{1} = y'_{1} - u(s_{1}) \simeq y'_{1}$ in virtue of (\ref{Eq:16}), and hence the Halpern functional integral reduces to a set of ordinary integrals, as explained in Section~\ref{sec:FradkinRepresentations}, and represented by the Bundle Diagram of Figure~{\ref{Fig:1}}.

Here, the time and longitudinal coordinates of the end-points of the bundle are the same, although their transverse coordinates -- measured vertically in the figure -- are separated.  Bundle diagrams are not Feynman diagrams, but are perhaps a more efficient way of representing the sum over all of the Feynman graphs corresponding to such multiple gluon exchange.

\begin{figure}
\includegraphics[keepaspectratio,height=35mm]{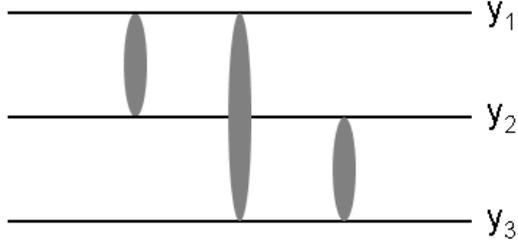}%
\caption{\label{Fig:2}Gluon bundles exchanged among three quarks.}
\end{figure}

A slightly more complicated expression describes gluon bundles exchanged between any two of three quarks, as in Figure~{\ref{Fig:2}}, where, because of EL, the $w$-coordinates of each of the $(f \cdot \chi)^{-1}$ entering into the appropriate Halpern functional integral are the same, even though the transverse coordinates of the three quarks can be quite different.

\begin{figure}
\includegraphics[keepaspectratio,height=35mm]{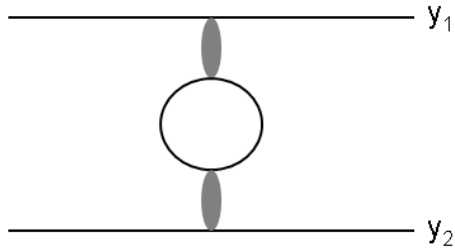}%
\caption{\label{Fig:3}Gluon bundles joining two quarks via a closed quark loop.}
\end{figure}

In contrast, were a closed quark loop -- corresponding to a simple relaxation of the quenched approximation -- to appear between a pair of quarks, joined to each external quark lines by the exchange of a gluon bundle, as in Figure~{\ref{Fig:3}}, there will now be two distinct sets of ordinary Halpern integrals to be evaluated.

As will be seen elsewhere, the effective diagram of Figure~{\ref{Fig:3}} will provide us with the essential features of the Nucleon-Nucleon potential for separation lengths beyond 2 fm~\cite{Fried2011_QCD5}.

\section{Summary}\label{sec:Summary}

For a whole family of QCD processes, a remarkable property of Effective Locality was first observed on an approximated (quenched and eikonal) version of QCD~\cite{Fried2009_QCD1}. However, that property still holds without these approximations, so that Effective Locality appears as a genuine property of non-perturbative QCD~\cite{Fried2010_QCD2}.

In the present article, a thorough analysis of the constraints inherent to Effective Locality has displayed an odd $\delta^{(2)}(\vec{b})$-term in the argument of an exponential factor describing $Q/\bar{Q}$ scattering at impact parameter $\vec{b}$. This fact is here taken as an indication that an untenable hypothesis is lurking in the derivation leading to this $\delta^{(2)}(\vec{b})$-function, an assumption most certainly inherited from our long practice of perturbative QCD.

Accordingly, a replacement of the meaningless $\delta^{(2)}(\vec{b})$ in an exponential factor of an exact statement obtained in an 'ideal' QCD amplitude, by a phenomenologically defined $\varphi(\vec{b})$ is proposed. It can be performed for each and every QCD process and will be used for the derivation of quark binding in a following article~\cite{Fried2011_QCD4}.  As emphasized in the text, the proposed transition from 'ideal' to 'realistic' QCD is phenomenological, but it clearly reflects the necessary change in viewpoint, taking into account the fact that quark quanta are always bound.

It is only when a complete sum over all relevant Feynman graphs has been accomplished in a gauge-invariant way, that the new and exact property of Effective Locality appears.  That property, if taken seriously, then forces this change to a would-be more 'realistic' form of non-perturbative QCD where transverse imprecision would be built-in.

Clearly, work remains to be done in order to appreciate as deeply as possible the necessity of such a change as specified by (\ref{Levy}). This point will be dealt with, in detail, in Ref.~\cite{Fried2011_QCD4}.  At face value, though, it is certainly fascinating that starting from a notion of propagation, such as implemented through the familiar 2-point function $\mathbf{G}_{c}(x, y|A)$, with its obvious perturbative content, one could be lead, in the bound non-perturbative context, to another mode of behavior, more appropriately described in terms of Levy flight distributions!

% If in two-column mode, this environment will change to single-column
% format so that long equations can be displayed. Use
%\begin{widetext}
%\end{widetext}


\appendix

\section{\label{AppA} On the independence on the point $w_1=y_1-u(s_1)$ in Section~\ref{sec:FradkinRepresentations}}

One may then ask what happens when the variables $y_{1} - u(s_{1})$ change, such that space-time point $w_{1}$ changes to a different point, say $w'_{1}$?  The answer is that nothing concerning the output of the integral $\int{\mathrm{d}[\chi]}$ is changed, for all the other space-time points $w \neq w'_{1}$ are then removed, with their normalizations, and the remaining integration of $\int{\mathrm{d}^{n} \chi(w'_{1})}$ has exactly the same form as that of $\int{\mathrm{d}^{n} \chi(w_{1})}$.  In brief, for this scattering process, there is but a single, relevant $\int{\mathrm{d}^{n} \chi(.)}$ whose origin is irrelevant.

One may also ask if a significant Jacobian of the transformation between $\chi(w_{1})$ and $\chi(w'_{1})$ could enter the integrand.  Holding the external position coordinates constant, and treating the $\chi(w)$ as continuous functions, could there be a significant Jacobian under a change of value of $u(s_{1})$?  The answer is negative, and for two reasons:
\begin{enumerate}
  \item Inspection of the Fradkin representation for $\mathbf{G}_{c}[A]$ shows that these $u_{\mu}$ are bounded, $|u_{\mu}| \leq \sqrt{s} \leq m^{-1}$, where $m$ denotes the quark mass, while $y$ is a macroscopic quantity, so that the difference of $y_{1}$ and $u(s_{1})$ is effectively just $y_{1}$.
  \item Thanks to (\ref{Eq:16}), $s_{1}$ is fixed at 0 and $u(0)=0$; any other possible value of $s_{1}$ is suppressed by the EL property.
\end{enumerate}

\noindent For these two reasons, we infer that any such Jacobian is essentially unity, and that the final integral is independent of the value of the argument $w$ of $\chi(w)$.

\section{\label{AppB} On the approximation $ y'_{\perp}\simeq y_{\perp}$ in Section~\ref{sec:RealisticQCD}}

Before transverse imprecision was introduced, EL had the effect of attaching to the representative symbol $[f \cdot \chi(w)]^{-1}$ of each gluon bundle -- exchanged between quark and/or antiquark of respective CM coordinates $y_{1}$ and $y_{2}$ -- a pair of delta functions, $\delta^{(4)}(w - y_{1} + u(s_{1})) \delta^{(4)}(y_{1} - y_{2} + \bar{u}(s_{2}) - u(s_{1}))$, as used in the text, or the pair $\delta^{(4)}(w - y_{1} + s_{1} p_{1}) \delta^{(4)}(y_{1} - y_{2} + s_{2} p_{2} - s_{1} p_{2})$ as used in an eikonal approximation~\cite{Fried2009_QCD1}.  For either case one finds fixed values of $w_{0}$ and $w_{\mathrm{L}}$, and $\vec{w}_{\perp} = \vec{y}_{1\perp} =- \vec{y}_{2\perp}$.  Then, as claimed in the text, the Halpern FI can be reduced to an ordinary set of $\int{\mathrm{d}^{n}\chi}$ integrals. In the process, though, one makes a systematic error, of the eikonal-type, by neglecting variations of the impact parameter or, correspondingly, of momentum transfer in the core parts of the matrix element.  In the context of the more exact expression of the first pair of delta functions above, that \emph{ad hoc} approximation avoided the much more complicated analysis of the transverse Fradkin's difference $u_{\perp}(s_{1}) - \bar{u}_{\perp}(s_{2})$.

With transverse imprecision now being included, the situation changes for the better, in the sense that no such approximation need be made.  But this change now requires a slightly more complicated justification of the argument which replaces Halpern's FI by a set of ordinary integrals.  For the question arises if this useful simplification is also true when the $\vec{w}_{\perp}$ inside the $[f \cdot \chi(w)]^{-1}$ factor is itself given by $\vec{y}^{\, \prime}_{\perp}$, and is being integrated over the $\int{\mathrm{d}^{2} \vec{y}^{\, \prime}}$ in that exponential factor, as in the discussion of the text leading to (\ref{Eq:20}).  It was there noted that the replacement of that $\vec{w}_{\perp}$ by $\vec{y}_{1\perp}$ or $-\vec{y}_{2\perp}$ is a reasonable approximation.
\par\medskip
The following argument is intended to give that approximation a more detailed justification.

Consider the Halpern FI
\begin{eqnarray}\label{Eq:B1}
\mathcal{N} \int{\mathrm{d}[\chi] \, [\det{(f \cdot \chi)}]^{-\frac{1}{2}} \, e^{\frac{i}{4}\int{\mathrm{d}^{4}w \, \chi^{2}} + ig\int{\mathrm{d}^{2}\vec{y}^{\, \prime}_{\perp} \, \mathfrak{a}(\vec{y}_{1\perp} - \vec{y}^{\, \prime}_{\perp}) \, \mathfrak{a}(\vec{y}_{2\perp} - \vec{y}^{\, \prime}_{\perp}) [f \cdot \chi(\vec{y}^{\, \prime}_{\perp})]^{-1}} } }
\end{eqnarray}
											
\noindent where $y'_{\mu} = (y_{0}; y_{\mathrm{L}}, \vec{y}^{\, \prime}_{\perp})$, the normalization is defined so that the FI of (\ref{Eq:B1}) equals  1 when $g = 0$. The dependence of color, time and longitudinal coordinate has been omitted for simplification of presentation.

As in the definition of this or any such FI, $\int{\mathrm{d}^{4}w \, \chi^{2} }$ is understood to mean $\delta^{4} \, \sum_{\ell=1}^{N}{\chi^{2}_{\ell}}$, where the subscript $\ell$ denotes the value of $\chi$ at the space-time point $w_{\perp \ell}$, and $\delta^{4}$ corresponds to a small volume surrounding that point, which is to become arbitrarily small as $N$ becomes arbitrarily large~\cite{Zee}.  As mentioned in the text by the end of Section~\ref{sec:RealisticQCD}, residual $\delta$-dependence will be re-expressed in terms of physically significant quantities as a last step; but for the following argument, all the transverse coordinate differences are to be taken as arbitrarily small.

Now, re-scale the $\chi_{\ell}$ variables such that $\delta^{2} \chi_{\ell} = \bar{\chi}_{\ell}$, and re-write (\ref{Eq:B1}) as
\begin{eqnarray}\label{Eq:B2}
\bar{\mathcal{N}} \int{\mathrm{d}[\bar{\chi}] \, [\det{(f \cdot \bar{\chi})}]^{-\frac{1}{2}} \, e^{\frac{i}{4} \, \sum_{\ell}{\bar{\chi}_{\ell}^{2}} + ig \delta^{2} \, \int{\mathrm{d}^{2}\vec{y}^{\, \prime}_{\perp} \, \mathfrak{a}(\vec{y}_{1\perp} - \vec{y}^{\, \prime}_{\perp}) \, \mathfrak{a}(\vec{y}_{2\perp} - \vec{y}^{\, \prime}_{\perp}) [f \cdot \bar{\chi}(\vec{y}^{\, \prime}_{\perp})]^{-1}} } }
\end{eqnarray}

Let us also break up the $\int{\mathrm{d}^{2}\vec{y}^{\, \prime}_{\perp}}$ integral into an infinite series of terms: One is free to choose the individual $\vec{y}^{\, \prime}_{\perp}$ coordinates as exactly those which define the transverse positions of the $\bar{\chi}_{\ell} = \bar{\chi}(w_{\ell})$.  In this way, (\ref{Eq:B2}) may be re-written as
\begin{eqnarray}\label{Eq:B3}
\bar{\mathcal{N}} \int{\mathrm{d}[\bar{\chi}] \, [\det{(f \cdot \bar{\chi})}]^{-\frac{1}{2}} \, e^{\frac{i}{4} \, \sum_{\ell}{\bar{\chi}_{\ell}^{2}} + ig \delta^{2} \, \Delta^{2}_{y_{\perp}} \, \sum_{\ell}{\mathfrak{a}(\vec{y}_{1\perp} - \vec{y}^{\, \prime}_{\perp\ell}) \, \mathfrak{a}(\vec{y}_{2\perp} - \vec{y}^{\, \prime}_{\perp \ell}) [f \cdot \bar{\chi}(\vec{y}^{\, \prime}_{\perp\ell})]^{-1}} } }
\end{eqnarray}

\noindent where $\Delta^{2}_{y_{\perp}}$ is understood as a true infinitesimal quantity, and where, for simplicity, we suppress explicit dependence on $y_{0}$ and $y_{\mathrm{L}}$.  But now (\ref{Eq:B3}) may be written as the product of $N$ integrals,
\begin{eqnarray}\label{Eq:B4}
&& \prod_{\ell}^N \bar{\mathcal{N}}_{\ell} \int{\mathrm{d}^{n}\bar{\chi}(\vec{y}^{\, \prime}_{\perp \ell}) \, [\det{(f \cdot \bar{\chi}(\vec{y}^{\, \prime}_{\perp \ell}))}]^{-\frac{1}{2}} \, e^{\frac{i}{4} \, {\bar{\chi}^{2}(\vec{y}^{\, \prime}_{\perp \ell})} + ig \delta^{2} \, \Delta^{2}_{y_{\perp}} \, {\mathfrak{a}(\vec{y}_{1\perp} - \vec{y}^{\, \prime}_{\perp\ell}) \, \mathfrak{a}(\vec{y}_{2\perp} - \vec{y}^{\, \prime}_{\perp \ell}) [f \cdot \bar{\chi}(\vec{y}^{\, \prime}_{\perp\ell})]^{-1}} } } \\ \nonumber &\equiv& \prod_{\ell}^N{\mathbb{F}(ig \delta^{2} \, \Delta^{2}_{y_{\perp}} \, \mathfrak{a}(\vec{y}_{1\perp} - \vec{y}^{\, \prime}_{\perp\ell}) \, \mathfrak{a}(\vec{y}_{2\perp} - \vec{y}^{\, \prime}_{\perp \ell}))} \\ \nonumber &\equiv& \prod_{\ell}^N{\mathbb{F}(z_{\ell})},
\end{eqnarray}
												
\noindent where (\ref{Eq:B4}) denotes the normalized product of all such $(\vec{y}^{\, \prime}_{\perp \ell})$-valued integrals, and $\mathbb{F}(z_{\ell})$ denotes the ordinary integral $\int{\mathrm{d}^{n}\bar{\chi}_{\ell}}$ over the variable associated with $\vec{y}^{\, \prime}_{\perp \ell}$.  That integral is well defined in the sense that, for $|z_{\ell}| < 1$, as is the case here, it can be expressed as an absolutely-convergent series, or as a converging integral over a set of eigenvalues in a random matrix calculation.

One then expects to be able to write $\mathbb{F}(z)$ in terms of its Fourier transform,
\begin{equation*}
\mathbb{F}(z_{\ell}) = \int{\mathrm{d}\varrho \, \tilde{\mathbb{F}}(\varrho) \, e^{i z_{\ell} \varrho}}
\end{equation*}

\noindent where the normalization condition of (\ref{Eq:B4}) stipulates that $\int{\mathrm{d}\varrho \, \tilde{\mathbb{F}}(\varrho)} = 1$.  Since $z_{\ell}$ is proportional to the infinitesimal $\Delta^{2}_{y_{\perp}}$, one may expand in powers of $z_{\ell}$,
\begin{equation*}
\mathbb{F}(z_{\ell}) = \int{\mathrm{d}\varrho \, \tilde{\mathbb{F}}(\varrho) \, \left[ 1 + i z_{\ell} \varrho + \cdots \right]} = 1 + i \int{\mathrm{d}\varrho \, \varrho \, \tilde{\mathbb{F}}(\varrho)} \, z_{\ell} + \cdots,
\end{equation*}

\noindent so that (\ref{Eq:B4}) becomes
\begin{eqnarray}\label{Eq:B5}
&& \prod_{\ell}{ \left[ 1 + i \int{\mathrm{d}\varrho \, \varrho \, \tilde{\mathbb{F}}(\varrho)} \, z_{\ell} \right]} \\ \nonumber &=& 1 + i \int{\mathrm{d}\varrho \, \varrho \, \tilde{\mathbb{F}}(\varrho)} \, \sum_{\ell}{z_{\ell}} \\ \nonumber &=& 1 + i \int{\mathrm{d}\varrho \, \varrho \, \tilde{\mathbb{F}}(\varrho)} \cdot ig  \delta^{2} \, \int{\mathrm{d}^{2}\vec{y}^{\, \prime}_{\perp} \, \mathfrak{a}(\vec{y}_{1\perp} - \vec{y}^{\, \prime}_{\perp}) \, \mathfrak{a}(\vec{y}_{2\perp} - \vec{y}^{\, \prime}_{\perp})}.
\end{eqnarray}

\noindent With
\begin{equation}
\int{\mathrm{d}^{2}\vec{y}^{\, \prime}_{\perp} \, \mathfrak{a}(\vec{y}_{1\perp} - \vec{y}^{\, \prime}_{\perp}) \, \mathfrak{a}(\vec{y}_{2\perp} - \vec{y}^{\, \prime}_{\perp})} = \int{\frac{\mathrm{d}^{2}q}{(2 \pi)^{2}}} \, \left| \tilde{\mathfrak{a}}(q) \right|^{2} \, e^{iq \cdot (\vec{y}_{1\perp} - \vec{y}_{2\perp})} \equiv \varphi(\vec{b}),
\end{equation}

\noindent (\ref{Eq:B4}) becomes
\begin{equation*}
\int{\mathrm{d}\varrho \, \tilde{\mathbb{F}}(\varrho) \, \left[ 1 + i \varrho (ig \delta^{2}) \varphi(\vec{b}) \right] },
\end{equation*}

\noindent which is just the first-order expansion of the result obtained in the text when $\vec{y}^{\, \prime}_{\perp}$ was shifted to $\vec{y}^{\, \prime}_{1\perp}$ or $-\vec{y}^{\, \prime}_{2\perp}$.  And since $\delta^{2} g \varphi$ is expected to be small, $\delta^{2} \varphi << 1$, it is, in effect, equivalent to
\begin{equation*}
\int{\mathrm{d}\varrho \, \tilde{\mathbb{F}}(\varrho) \, e^{i \varrho (ig \delta^{2}) \varphi(\vec{b})} },
\end{equation*}

\noindent which is just the integral of (\ref{Eq:B2}) when the intuitively equivalent change $\vec{y}^{\, \prime}_{\perp} \rightarrow \vec{y}^{\, \prime}_{1\perp}$ has been made in the argument of $(f \cdot \bar{\chi})^{-1}$, and after the residual $\delta^{2}$ dependence has been continued to the measurably-significant value of $1/(E  \mu)$.

%\begin{acknowledgments}
%One of us (H. M. Fried) would like to thank the Julian Schwinger Foundation for the Travel Grant.
%\end{acknowledgments}


\begin{thebibliography}{}
\bibitem{Fried2009_QCD1}
H. M. Fried, Y. Gabellini, T. Grandou and Y.-M. Sheu, Eur. Phys. J. \textbf{C65}, (2010) 395.

\bibitem{Fried2010_QCD2}
H. M. Fried, M. Gattobigio, T. Grandou and Y.-M. Sheu, arXiv:1003.2936 [hep-th].

%\bibitem{Fried2011_QCD3}
%H. M. Fried, T. Grandou and Y.-M. Sheu, arXiv:1103.4179 [hep-th].

\bibitem{Fried2011_QCD4}
H. M. Fried, Y. Gabellini, T. Grandou and Y.-M. Sheu, arXiv:1104.4663 [hep-th].

\bibitem{Fried2011_QCD5}
H. M. Fried, Y. Gabellini, T. Grandou and Y.-M. Sheu, in preparation.

\bibitem{Kouneiher2010}
J. Kouneiher, private commninication.

\bibitem{Fradkin1966}
E. S. Fradkin, Nucl. Phys. \textbf{76}, (1966) 588.

\bibitem{Halpern1977a}
M. B. Halpern, Phys. Rev. D\textbf{16}, (1977) 1798.

\bibitem{Halpern1977b}
M. B. Halpern, Phys. Rev. D\textbf{16}, (1977) 3515.

\bibitem{levy}
For example, see \url{http://en.wikipedia.org/wiki/Stable_distribution}


\bibitem{HMF1}
H. M. Fried, \textit{Functional Methods and Models in Quantum Field Theory} (The MIT Press, Cambridge, MA 1972)

\bibitem{HMF2}
H. M. Fried, \textit{Basics of Functional Methods and Eikonal Models} (Editions Fronti\`{e}res, Gif-sur-Yvette Cedex, France 1990)

\bibitem{HMF3}
H. M. Fried, \textit{Green's Functions and Ordered Exponentials} (Cambridge University Press, Cambridge 2002)

\bibitem{Lapidus2000}
G. W. Johnson and M. L. Lapidus, \textit{The Feynman Integral and Feynman's Operational Calculus} (Oxford University Press, Oxford 2000)

\bibitem{GR}
I. S. Gradshteyn and I. M. Ryzhik, \textit{Table of Integrals, Series and Products} (Academic Press, London 1994) formula 8.253.1

\bibitem {Zee}
A. Zee, \textit{Quantum Field Theory in a Nutshell, 2nd Ed.} (Princeton University Press, Princeton 2010)

\bibitem{Schiff}
L. I. Schiff, \textit{Quantum Mechanics, 3rd Ed.} (McGraw-Hill, 1968)
% etc
\end{thebibliography}


\end{document}